\documentclass[twoside]{ilcws07}
\usepackage[latin1]{inputenc}
\usepackage[dvips]{graphicx,epsfig,color}
\usepackage{wrapfig,rotating}
\usepackage{amssymb,amsmath,array}

\begin{document}
\title{
Anomalous Gamma Gamma Interaction} 
\author{Philip Yock
\vspace{.3cm}\\
University of Auckland - Department of Physics \\
Auckland - New Zealand
}

\maketitle

\begin{abstract}
Data from LEP2 on hadron production in $\gamma\gamma$ interactions at high $p_T$ exceed the predictions of QCD by about an order of magnitude. The amplitude for the process is asymptotically proportional to the sum of the squares of the charges of quarks. The data are suggestive of models where quarks have unit charges, or larger. Unequivocal tests could be made with the ILC or CLIC, but a plasma wakefield $e^-e^-$ collider might provide the most affordable option~\cite{yock1}.   
\end{abstract}

\section{The anomaly}

\begin{wrapfigure}{r}{0.4\columnwidth}
\centerline{\includegraphics[width=0.3\columnwidth]{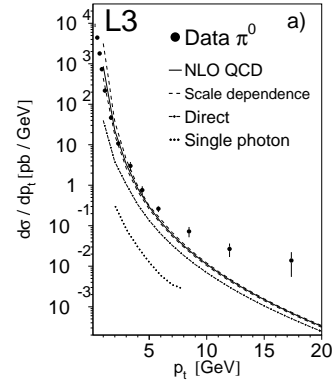}}
\caption{Differential cross-section for inclusive $\pi^0$ production in $\gamma\gamma$ interactions measured at LEP2 and the prediction by QCD} \label{Fig:1}
\end{wrapfigure}

Hints of anomalous hadron production in $\gamma\gamma$ interactions were first reported from studies made at SLAC and DESY~\cite{binnewies}. The first evidence for a large anomaly was reported by the L3 group from studies carried out at LEP2 at 200 GeV~\cite{achard}. L3 reported data on inclusive $\pi^0$ production at values of $p_T$ approaching 20 GeV/c. The data, shown in Figure~\ref{Fig:1}, exceed the QCD prediction by about an order of magnitude at the highest transverse momenta.  

The $\gamma\gamma$ interactions observed by L3 occurred in the process $e^-e^+\rightarrow e^-e^+ \pi^0 X$ where $\gamma's$ were radiated by the incoming $e^-$ and $e^+$. The radiative process was known to be a copious source of $\gamma\gamma$ interactions~\cite{brodsky}. Events were selected by L3 in which the outgoing $e^-$ and $e^+$ emerged in the beam pipe of the collider and, although undetected, were known to produce almost real ${\gamma}$'s. The ${\gamma\gamma}$ events were identified by calorimeter measurements in which the total energy of emerging hadrons was significantly less than 200 GeV.

Observations of $\pi^0$ production at LEP2 have not been reported by other groups. However, data on $\pi^\pm$ production have been reported by both L3 and OPAL~\cite{achard,wengler}. They are comparable to the $\pi^0$ data. Both L3 and OPAL found significant excesses over the QCD prediction at $p_T \sim 17$ GeV/c. The excess reported by L3 is slightly greater than that shown in Figure~\ref{Fig:1} for $\pi^0$ production, whereas for OPAL the excess is slightly less. 

Data have also been reported by L3 and OPAL on inclusive jet production~\cite{achard,wengler}. The data by L3 extend to higher values of $p_T$ than those of OPAL, and they show a greater excess over the QCD prediction, reaching nearly an order of magnitude at 45 GeV/c. Most of the above datasets are displayed in~\cite{yock1}.    

\section{Experimental remarks}
The main source of background in the above experiments results from $e^-e^+ \rightarrow Z\gamma$ where initial state radiation reduces the invariant mass of the $e^-e^+$ system to that of the Z, and the $\gamma$ emerges in the beam pipe and is undetected. Hadronic decays of the Z in these events can  mimic $\gamma\gamma$ interactions. However, Z's produced as above are boosted to an energy of $\sim0.65\sqrt{s}$. Both L3 and OPAL made cuts on the total energy observed in their calorimeters to remove the background. The high-precision calorimeters employed by L3 enabled them to place the cut at $0.4\sqrt{s}$ and maintain the background below 1\%~\cite{yock1}. In the case of OPAL, the cut was placed at a lower level, and/or extra conditions applied, resulting in analyses of greater complexity. 

\section{Physics implications}

\begin{wrapfigure}{r}{0.3\columnwidth}
\centerline{\includegraphics[width=0.25\columnwidth]{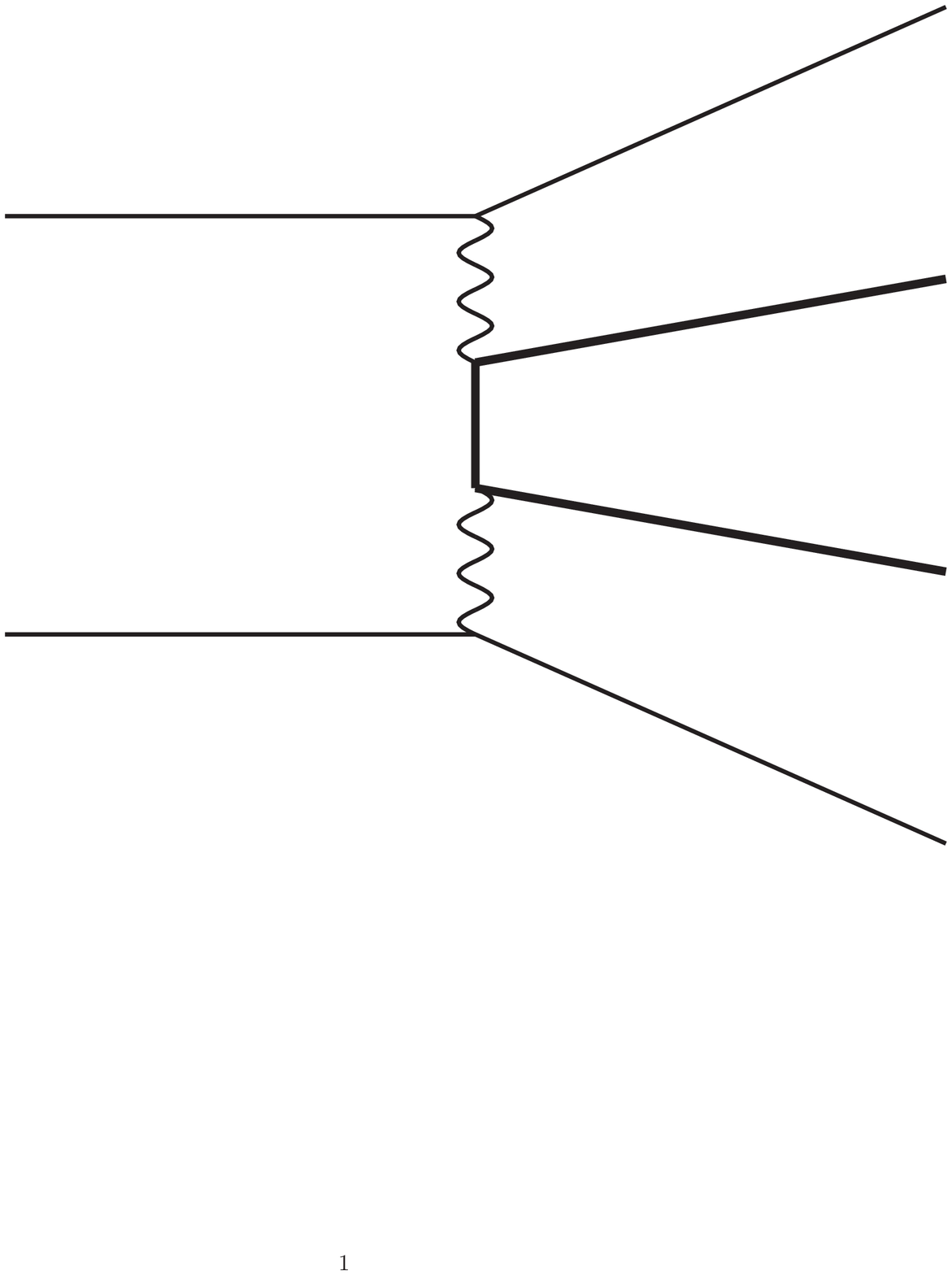}}
\caption{Feynman diagram for hadron production at asymptotically high $p_T$ in $\gamma\gamma$ interactions as observed with $e^-e^+$ colliders. Thin lines denote leptons, heavy lines quarks.} 
\label{Fig:2}
\end{wrapfigure}

The measurements by L3 and OPAL are significant. According to the Standard Model (SM), the Feynman diagram shown in Figure~\ref{Fig:2} applies when $p_T \geq$ few GeV/c~\cite{binnewies}. The amplitude was therefore predicted to be proportional to the sum of the squares of the charges of quarks accessible at 200 GeV, and the cross section expected to provide a simple but sensitive constraint on these basic charges~\cite{witten}. The data are, however, inconsistent with the SM expectation. If valid, they suggest alternative models with larger charges. 

\subsection{Unit charges} 
Ferreira~\cite{ferreira} considered a Han-Nambu model with unit charges and found improved agreement with the data, although the $p_T$ dependence was not perfectly fitted. He reviewed the extensive literature that exists on quarks of unit charges, and considered the production of b hadrons in $\gamma\gamma$ interactions where a possible excess of data over theory was reported. He found this excess could be accounted for by unit charges. We note, however, that the ALEPH group subsequently reported a lower cross section for the production of b hadrons~\cite{aleph}. 

\subsection{High charges}

Before QCD was proposed, two theories of highly charged quarks were proposed, independently. In one, quarks were proposed by Schwinger~\cite{schwinger} to be magnetic monopoles, in the other they were proposed by the author~\cite{yock2} to be highly electrically charged particles. In both cases it was assumed that strong electromagnetic forces between quarks accounted for quark binding, and that strong nuclear interactions between composite states occurred as residual interactions. Colour charges were not included. Both theories were based on eigenvalue equations for charge. In~\cite{schwinger} a modified version of Dirac's eigenvalue equation for magnetic charge was assumed. In~\cite{yock2} a physical solution to a Gell-Mann-Low equation was assumed. In both theories, little progress was made in solving the field equations, because of the strong couplings involved. 

\begin{wrapfigure}{r}{0.4\columnwidth}
\centerline{\includegraphics[width=0.35\columnwidth]{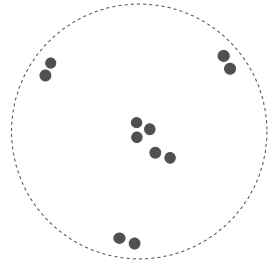}}
\caption{Generalized Yukawa model of the nucleon. The dots represent highly charged quarks clustered to form bare mesons and nucleons. The larger scale is the $\pi$ Compton wavelength, i.e. the observed size of the nucleon. The smaller scale represents the fundamental separation between quarks in bound states, assumed in~\cite{yock3} to be $\sim1,000\times$ smaller.} 
\label{Fig:3}
\end{wrapfigure}

A generalized Yukawa model~\cite{yock3} was proposed for~\cite{yock2}, as shown in Figure~\ref{Fig:3}. Bare mesons and nucleons were assumed to act as partons of unit charge at low energies, thus avoiding the hadronization puzzle of the SM. Large electromagnetic effects were anticipated to occur at the shorter scale shown in the figure. The possibility arises that the LEP2 data are exhibiting the onset of large electromagnetic effects, because the $\gamma\gamma$ amplitude is proportional to the square of the charges of the participating particles. Measurements at higher energies would provide a clear test of this model.    

The data shown in Figure~\ref{Fig:1} are similar to Rutherford's old $\alpha$-particle scattering data, here reproduced in~\cite{yock1}. Both cross-sections fall in similar fashion through $\sim5.5$ orders of magnitude. It was, of course, Rutherford's data that supplanted J.J.~Thomson's model of the atom in which low-mass electrons swarmed in a cloud of massless, positive charge, a model that bears some resemblance to today's SM of the nucleon. In contrast, Rutherford's nuclear atom bears more resemblance to particle models with highly charged quarks. We note that the latter require multiple generations of quarks that are not replications of one another~\cite{schwinger,yock2}. They thus possess inbuilt symmetry breaking. Current searches for the Higgs boson may therefore be viewed as null tests of these models~\cite{yock4}.  
                
\section{Future experiments}

As remarked above, the $\gamma\gamma$ interaction probes fundamental parameters of the SM, sensitively. Existing data from LEP2 by L3 and OPAL, whilst in approximate agreement with one other, are in serious disagreement with the SM. It would clearly be useful if the remaining LEP2 groups (ALEPH and DELPHI) subjected their data to similar analyses. Studies of $\gamma\gamma$ interactions at higher energies would, however, be of greater interest, because neither the L3 nor the OPAL datasets reached asymptotic limits.   

The International Linear Collider (ILC) and the Compact Linear Collider (CLIC) would both be ideal machines for this purpose. Plans for the former are more advanced, but both could be used to study $\gamma\gamma$ interactions. Purpose-built $\gamma\gamma$ colliders would not be required. This has been amply demonstrated by the L3 and OPAL experiments. Moreover, both ILC and CLIC could be operated as $e^-e^-$ colliders. In this mode, background from $e^-e^+$ annihilation would be absent. The cuts on $\sqrt{s}$ described in Section 2 would not be required, and the full beam energy of the colliders could be utilised, thereby accessing significantly higher values of $p_T\sim100$ GeV/c given sufficient luminosity. 
     
A plasma wakefield collider~\cite{joshi1} could offer a more affordable means for carrying out the above program, especially if the Higgs boson fails to materialize at Fermilab and at the LHC. Both electrons and positrons can be accelerated in plasma wakefields. However, an $e^-e^-$ configuration would greatly reduce the cost. Moreover, as remarked above, it would be subject to greatly reduced background. 

\section{Conclusions} 

Despite often-made claims to the contrary, e.g.~\cite{nature}, the SM has not withstood every challenge. For example, Yukawa's meson theory, confirmed by the discovery of the $\pi$-meson in 1947, appears to lie outside the model\footnote{In~\cite{derrick} a pion term was clearly required but added 'by hand'.}. Similarly, the lifetimes of the strange particles, also discovered in 1947, lie outside the model~\cite{bass}. A  raison d'\^{e}tre for the strange particles has yet to be provided, and, conversely, glueballs have yet to be confirmed. The $\gamma\gamma$ results present additional challenges. They are suggestive of alternative models with quarks of unit or larger charges, but further data at higher energies are needed. Three machines would be capable of providing such data - ILC, CLIC, and a plasma wakefield collider. Of these, plans for the first are most advanced, but the last could eventually prove most affordable.  

\section*{Acknowledgements}
Pablo Achard, Barry Barish, Bob Bingham, John Campbell, Francis Collins, Pedro Ferreira, Brian Foster, Philippe Gavillet, Chan Joshi, Maria Kienzle-Focacci, Paolo Palazzi, Francois Richard, Valery Telnov and Thorsten Wengler are thanked for correspondence.


\begin{footnotesize}



%

\end{footnotesize}


\end{document}